\documentclass[a4paper,aps,pra,twocolumn]{revtex4-2}

\usepackage[english]{babel}
\usepackage[utf8x]{inputenc}
\usepackage{amsmath}
\usepackage{graphicx}
\usepackage{mathtools}
\usepackage{bm}
\usepackage{xcolor}
\usepackage{float}
\makeatletter
\let\newfloat\newfloat@ltx
\makeatother
\usepackage{algorithm}
\usepackage{algpseudocode}

\begin{document}

\title{Curve-Fitted QPE: Extending Quantum Phase Estimation Results for a Higher Precision using Classical Post-Processing}
\author{See Min Lim$^{1}$}
\email[]{seemin.lim@u.nus.edu}

\author{Cristian E. Susa$^{2}$}
\email[]{cristiansusa@correo.unicordoba.edu.co}

\author{Ron Cohen$^{3}$}
\email[]{ron@classiq.io}

\affiliation{$^1$Department of Physics, National University of Singapore, Singapore}
\affiliation{$^2$Department of Physics and Electronics, University of C\'ordoba, 230002 Monter\'ia, Colombia}
\affiliation{$^3$Classiq Technologies, Tel Aviv, Israel }

\begin{abstract}
    Quantum Phase Estimation (QPE) is a crucial component of several front-running quantum algorithms. Improving the efficiency and accuracy of QPE is currently a very active field of research. In this work, we present a hybrid quantum-classical approach that consists of the standard QPE circuit and classical post-processing using curve-fitting, where special attention is given to the latter. We show that our approach achieves high precision with optimal Cramér–Rao lower bound performance and is comparable in error resolution with the Variational Quantum Eigensolver and Maximum Likelihood Amplitude Estimation algorithms. Our method could potentially be further extended to the case of estimating multiple phases. 
\end{abstract}

\maketitle
\begin{sloppypar}

\section{Introduction} \label{section:introduction}

Quantum phase estimation (QPE) is an important building block for developments in quantum computing, playing a critical role in algorithms that show strong evidence for quantum advantage, such as Shor's Algorithm \cite{shor1994algorithms} for prime factorization and the Harrow–Hassidim–Lloyd Algorithm \cite{harrow2009quantum} for solving systems of linear equations. Given the progress of early fault-tolerant quantum computing \cite{hongkang2023}, experimental demonstration \cite{yamamoto2024}, and the adaptability to more general phase estimation protocols \cite{papadopoulos2024}, amongst other developments, the QPE circuit is particularly worth attention.

However, there are both near-term and long-term issues with the QPE algorithm due to current and permanent hardware constraints. The traditional QPE algorithm \cite{kitaev1995quantum} requires a very large number of qubits, large depth, and long coherence time to achieve good precision. In the current noisy intermediate-scale quantum (NISQ) era, these requirements pose a significant obstacle to near-time use due to the lack of fault-tolerance \cite{mohammadbagherpoor2019improved}.

More permanently, QPE has the inherent issue of spectral leakage. Spectral leakage refers to the measurement result leaking from the exact result \cite{sakuma2024entanglement}, which will occur when the number of qubits cannot capture the accurate phase precisely \cite{xiong2022dual}. Since an irrational phase can never be entirely captured by the recording qubits, there is always an intrinsic risk of spectral leakage, reducing the accuracy and efficiency of the QPE algorithm.

To mitigate these constraints to attain a higher accuracy effectively, we propose an alternative hybrid algorithm Curve-Fitted QPE. As in the standard QPE algorithm, we will execute the QPE circuit multiple times to find the probability distribution. However, instead of utilizing only the result with the highest probability, our method applies curve-fitting across the whole distribution, which allows us to achieve a more precise estimate than traditionally possible with a given number of recording qubits. Effectively, our method extends the quantum results to a higher precision by accounting for quantum information (specifically, the probability distribution of all states), which is usually discarded in the traditional QPE algorithm.

Other hybrid alternatives such as the Variational Quantum Eigensolver (VQE) and Maximum Likelihood Amplitude Estimation (MLAE) have been proposed \cite{suzuki2020amplitude,peruzzo2014variational,grinko2021iterative}. The latter method MLAE similarly considers maximising likelihood but it attempts to estimate the amplitude while our method estimates the phase. \cite{cruz2020optimizing} also uses the least squares regression method to fit the phase estimates, however their work focuses on the time evolution operator and estimates the phase over time. Our work is also similar to \cite{o2019quantum, lu2022unbiased} in that classical post-processing is done. Our method can improve the QPE precision using only an additional curve-fitting process, without the need for additional quantum resources. We will demonstrate in this paper that our method has optimal Cramér–Rao Lower Bound (CRLB) performance, and is comparable in error resolution with the VQE and MLAE algorithms. 

This paper is organized as follows. In Section \ref{section:method} we briefly provide an overview of our Curve-Fitted QPE algorithm, then derive the corresponding Cramér–Rao lower bound for the the QPE algorithm's probability function and present in detail our curve-fitting implementation. In Section \ref{section:results} we present our main results, discussing the accuracy and error resolution of our method against other approaches as VQE. Finally, Section \ref{section:conclusion} concludes with further elaboration on the utility of our method, as well as possible applications and further work.

\section{Theory and Method}\label{section:method}

Here, we give an overview of our method. First, we analytically compute the probability mass function of the QPE, which is also given in \cite{nielsen2010quantum}. Following the mathematics of the original QPE algorithm, the probability mass function is given by (see Appendix \ref{app:pmf}) for details:

\begingroup \small\begin{equation} 
P(y)  = \frac{1}{M^2} \left( \frac{ 1 - \cos(2\pi\left(y - \theta M \right))}{ 1 - \cos(\frac{2\pi }{M}\left(y - \theta M \right))} \right)
\label{eqn:pmf}
\end{equation} \endgroup

This equation \ref{eqn:pmf} gives the probability of measuring the state \(y\) on a QPE with \(n\) recording qubits where \(M = 2^n\), when the initial state of the system is \(\vert 0 \rangle^n \otimes \vert \psi \rangle\). \(\vert \psi \rangle\) is an eigenvector of the QPE's unitary with the eigenvalue \(e^{2\pi i \theta}\). 

\begin{figure}[!h]
    \begin{center}
    \setkeys{Gin}{width=\linewidth}
    \includegraphics{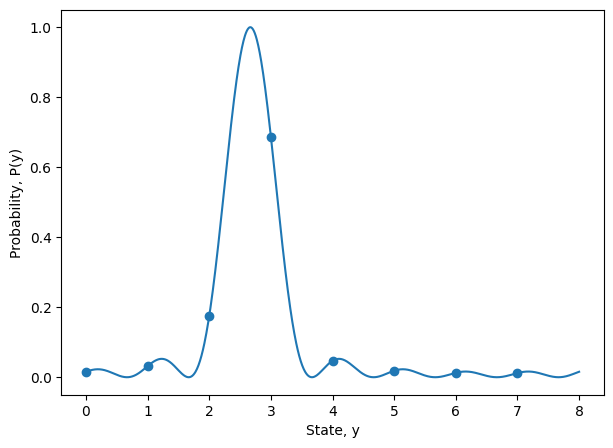}
    \end{center}
    \caption{Curve-fitted QPE results with 1000000 shots and 3 recording qubits. The actual phase was \(\frac{1}{3}\) and the estimated phase was \(0.33331287\). This is an improvement over the traditional QPE estimated phase of \(\frac{3}{8}\) or \(0.375\).} 
    \label{fig:example}
\end{figure}

Our method then curve-fits this probability equation to the distribution results of multiple executions of the QPE algorithm, from which we will attain our phase estimation. Figure \ref{fig:example} shows an example of the curve-fitting on the QPE results histogram.

We will focus on the probability function in the following subsection to analyze the QPE algorithm error and detail our curve-fit implementation after in subsection \ref{subsection:curve_fitting}.

\subsection{Fisher Information and CRLB}\label{subsection:crlb}

From the probability mass function, the Cramér–Rao Lower Bound (CRLB) \cite{rice2007mathematical} for the probability mass function can be found, which gives us the fundamental error limit of the QPE algorithm \cite{liu2020quantum, yu2022quantum, fiderer2021general}, providing a benchmark to compare with our method's error range.

First, we can derive \(score(y)\), which indicates the sensitivity of the QPE algorithm to changes in \(\theta\) at a particular \(\theta\) \cite{ly2017tutorial, xiong2022dual, petz2011introduction}:

\begingroup \small \begin{align} 
    score(y) \coloneqq \frac{\partial \log{P(y)}}{\partial \theta} \nonumber & 2 \pi \left[ \frac{\sin{\frac{2\pi}{M}(y-\theta M)}}{1-\cos{\frac{2\pi}{M}(y-\theta M)}} \right.\nonumber\\
    &- \left.M \frac{\sin{2\pi}(y-\theta M)}{1 - \cos{2 \pi}(y-\theta M)}\right]
    \label{eqn:score}
\end{align} \endgroup

The Fisher Information (FI) of a single sample of the algorithm, which indicates the overall sensitivity of the QPE algorithm to changes in \(\theta\) is computed as \cite{ly2017tutorial, xiong2022dual, petz2011introduction}:
\begingroup \small
\begin{equation}
\begin{split}
FI & \coloneqq \sum_{y=0}^{M-1} (score(y))^2 P(y) \\
& = \frac{4 \pi ^2}{M^2} \sum_{y=0}^{M-1} \left[ \frac{\sin^2 \frac{2\pi}{M} (y-\theta M)(1-\cos{2\pi(y-\theta M)})}{(1-\cos{\frac{2\pi}{M}(y-\theta M)})^3}\right. \\
& + M^2 \frac{\sin^2{2\pi}(y-\theta M)}{(1-\cos{2\pi(y-\theta M)})(1-\cos{\frac{2\pi}{M}(y-\theta M)})} \\
& - 2M \left.\frac{\sin{\frac{2\pi}{M}(y-\theta M)} \sin{2\pi (y-\theta M)}}{(1-\cos{\frac{2\pi}{M}(y-\theta M)})^2} \right]\\
\end{split}
\end{equation}
\endgroup

When \((y - \theta M) \neq 0\) for all \(y\) for a given \(M\), our FI is independent of \(\theta\) and depends only on \(M\). This is the same condition for spectral leakage to occur,; hence, we are only focusing on this case. Based on this equation, we have included the Fisher Information for a single sample for each choice of M in Appendix \ref{app:fisher}.

The Fisher Information for the total number of shots, \(k\), is given by:
\begingroup \small
\begin{equation}
FI_{total} = k \cdot FI_{individual}
\end{equation}
\endgroup

The CRLB is given by 
\begingroup \small
\begin{equation}
MSE \geq \frac{1}{FI_{total}}
\end{equation}
\endgroup
which means that \(\frac{1}{FI_{total}}\) is the lowest possible Mean Squared Error.

\subsection{Curve-Fitting Method} \label{subsection:curve_fitting}

When \((y - \theta M) \neq 0\) for all \(y\) for a given \(M\), spectral leakage occurs and this is a permanent constraint of the QPE algorithm as discussed in the Section \ref{section:introduction}. At this stage, we add our main contribution involving classical post-processing to mitigate this limitation. For post-processing, we had specifically used SciPy's \verb|scipy.optimize.curve_fit| function \cite{2020SciPy-NMeth,vugrin2007confidence} to fit the probability mass function (equation \ref{eqn:pmf}) to our QPE results. This is a Non-Linear Least Squares problem, and the Trust Region Reflective algorithm was chosen for optimization. The notion of curve-fitting in our method provides a convenient and elegant visualization; statistically, our method is equivalent to optimizing for the parameter \(\theta\) of the generic probability mass function given the observed QPE results.

\begin{algorithm}
\caption{Post-Processing Algorithm} \label{algo:curve_fit}
\begin{algorithmic}[1]
\Require $n$, states, probs
\Ensure parameter, pcov
\State $M \gets 2^n$
\State guess $\gets$ states[probs.index(max(probs))]
\State left\_bound $\gets \frac{\text{guess} - 0.5}{M}$
\State right\_bound $\gets \frac{\text{guess} + 0.5}{M}$
\State bounds $\gets$ (left\_bound, right\_bound)
\Procedure{ProbabilityFunction}{$y, \theta$}
    \State Find probability of observing $y$ given $\theta$
    \State \textbf{return} $P(y)$
\EndProcedure
\State estimate1, variance1 $\gets$ curve\_fit(
ProbabilityFunction, states,
 probs, bounds, left\_bound) 
\State estimate2, variance2 $\gets$ curve\_fit(
ProbabilityFunction, states, probs, bounds, right\_bound)
\State \textbf{if} variance1 $<$ variance2 \textbf{then}
    \State \hspace{\algorithmicindent} \textbf{return} estimate1, variance1
\State \textbf{else}
    \State \hspace{\algorithmicindent} \textbf{return} estimate2, variance2
\end{algorithmic}
\end{algorithm}

Algorithm \ref{algo:curve_fit} describes the post-processing algorithm used. Given the QPE probability results in the array \verb|probs|, corresponding to each value in the array \verb|states| by index, we can find the traditional QPE result \verb|guess| which is the state \(y\) corresponding to the highest probability based on the QPE results. This gives the traditional QPE result of \(\theta_{\text{est}} = \frac{\text{guess}}{M}\), and it is clear that the real phase has to be within \(\pm \frac{0.5}{M}\) from this estimate, hence the \verb|left_bound|, \verb|right_bound| and \verb|bounds| can be found. We then define the function \verb|ProbabilityFunction| which takes the variables \(y\) and \(\theta\), returning the \(P(y)\) based on the probability mass function. To reduce the chances of returning local minima (especially when extending our method to a more general multi-phase case), two curve-fitting attempts starting from the \verb|left_bound| and \verb|right_bound| were made. Each curve-fitting attempt considers the \verb|ProbabilityFunction| which will be fitted onto the data, the QPE results \verb|states| and \verb|probs|, the boundaries for \(\theta_{\text{est}}\) \verb|bounds| and the starting point \verb|left_bound| or \verb|right_bound|. The \verb|ProbabilityFunction| can repeatedly generate the probabilities for each state in \verb|states| for a given \(\theta\), thus the \verb|curve_fit| function utilizes this to optimize over \(\theta\) to find a probability distribution best matching the observed \verb|probs|. Each curve-fitting attempt returns an estimate for \(\theta\) and the variance between the corresponding probabilities and the observed probabilities \verb|probs|. Our post-processing algorithm returns the set of \verb|estimate| and \verb|variance| which gives a lower variance. The returned \verb|estimate| is our algorithm's estimated phase \(\theta_{\text{est}}\). 

\section{Results and Discussion}\label{section:results}

In this section, we will conduct an error analysis of our method. We executed the Quantum Phase Estimation using 2 to 8 recording qubits and 10 to 1000000 shots (or executions of the circuit), through the Classiq platform \cite{Classiq}. We tested 4 known phases (\(\frac{1}{3}\), \(\frac{1}{5}\), \(\frac{1}{7}\) and \(\frac{1}{9}\)) separately, using a phase gate for the unitary. For each setting of \(n\) recording qubits, \(k\) number of shots and phase \(\theta\) for the phase gate, the experiment was repeated 100 times. Therefore, each plotted point in the diagrams in this section is created from 400 data points.

Here, we make a side note on the interpretation of \(M = 2^n\) as an indicator of circuit depth. From the QPE circuit, it is clear that the depth is proportional to the number of applications of the unitary, which is given by

\begingroup \small
\begin{equation*}
    2^{n - 1} + 2^{n - 2} +...+ 2^{0} = 2^n - 1.
\end{equation*}
\endgroup

Noting that the layer of Hadamard gates provides an additional circuit depth of 1 and that the inverse Quantum Fourier Transform is the hermitian conjugate of the Quantum Fourier Transform which has better than \(\mathcal{O}(2^n)\) scaling \cite{coppersmith2002approximate, hales2002quantum, cleve2000fast}, we can see that the depth scales according to \(M\).

\subsection{Comparison with traditional Quantum Phase Estimation}

\begin{figure}[!h]
    \begin{center}
    \setkeys{Gin}{width=\linewidth}
    \includegraphics{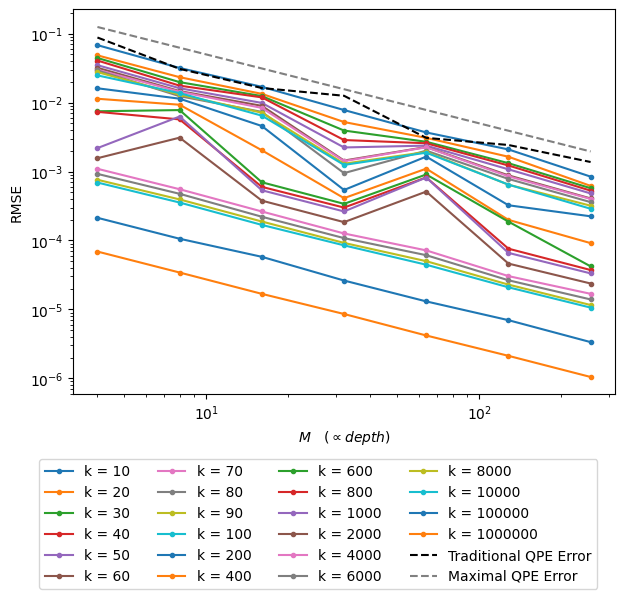}
    \end{center}
    \caption{RMSE against \(M\). Assuming the closest estimate is always found, the traditional QPE error for the 4 tested phases and the maximal error possible for a given M are overlaid.}
    \label{fig:compare_traditional}
\end{figure}

We can first compare our results with that of the traditional QPE error, based on the Root Mean Squared-Error (RMSE). We can see from Figure \ref{fig:compare_traditional} that our method consistently achieves the same or better error resolutions even with as little as \(10\) or \(20\) shots. Given that the traditional QPE method also requires multiple shots to ensure the closest estimate is found, our results suggest that our method can extend the precision without involving more quantum resources.
\begin{figure}[h]
    \begin{center}
    \setkeys{Gin}{width=\linewidth}
    \includegraphics{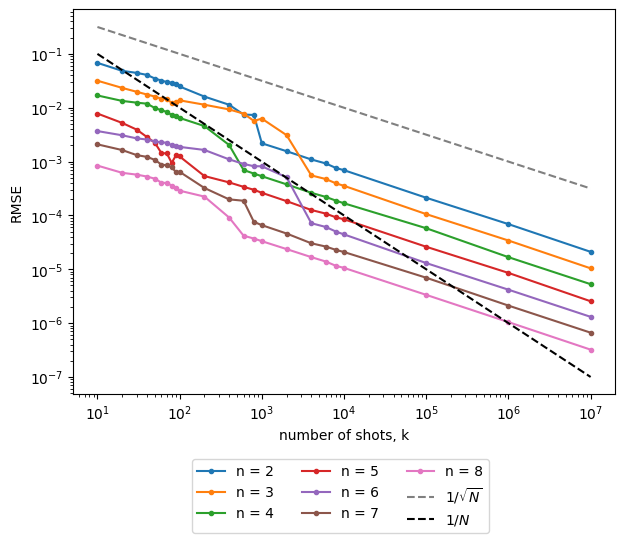}
    \end{center}
    \caption{RMSE against \(k\)}
    \label{fig:against_num_shots}
\end{figure}

\subsection{Order of Error Resolution}\label{subsection:resolution_order}

\begin{figure}[h]
    \begin{center}
    \setkeys{Gin}{width=\linewidth}
    \includegraphics{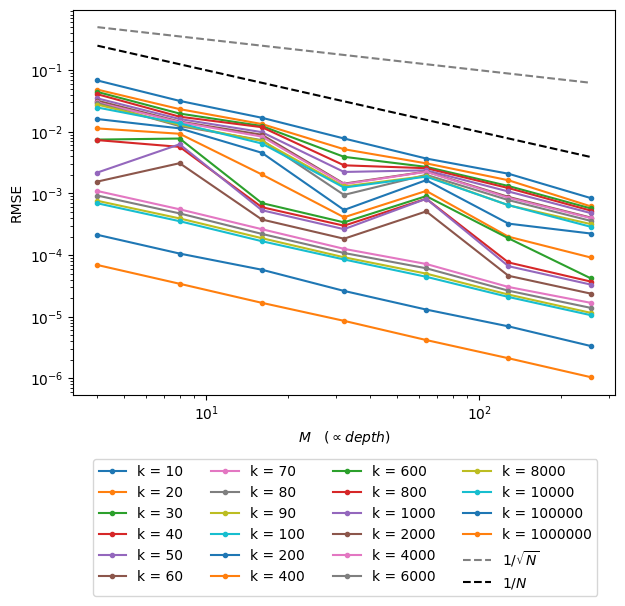}
    \end{center}
    \caption{RMSE against \(M\)}
    \label{fig:against_depth}
\end{figure}

We can further compare the order of our method's error with regards to \(k\) and depth separately in Figures \ref{fig:against_num_shots} and \ref{fig:against_depth}, which show that our method follows the Heisenberg's Limit \cite{xiong2022dual, gorecki2020pi} in terms of achieving \(\mathcal{O}(\frac{1}{\sqrt{k}M})\) scaling, an advantage over the Standard Quantum Limit of scaling on \(\mathcal{O}(\frac{1}{\sqrt{kM}})\) when considering the order of \(M\).

However, our method requires a reasonably large number of shots, therefore the predominant factor is \(k\) and our algorithm's RMSE effectively scales on \(\mathcal{O}(\frac{1}{\sqrt{k}})\). In other words, the number of shots scales on \(\mathcal{O}(\frac{1}{\epsilon^2})\) where \(\epsilon\) is the desired precision. This is the same scaling as in the case of VQE \cite{fedorov2022vqe}.

\subsection{Comparison with other NISQ-era algorithms}

Our estimation error scaling of \(\mathcal{O}(\frac{1}{\sqrt{k}M})\) meets the estimation error scaling of VQE and MLAE. However, our method is more straightforward and we will show that it can achieve optimal CRLB performance (i.e., optimal average absolute error) in the following subsection \ref{subsection:crlb_comparison}. 

Our method could extend to multiple phases, as shown in figure \ref{fig:examples_2phases}. The general probability equation is also derived in Appendix \ref{app:generic_pmf}. The guessed \(\theta\)s were \(0.49353036\) and \(0.33320002\). This is a potential advantage over the traditional QPE as our method can also differentiate the effects of two different eigenvalues to some extent.

\begin{figure}[h]
    \begin{center}
    \setkeys{Gin}{width=\linewidth}
    \includegraphics{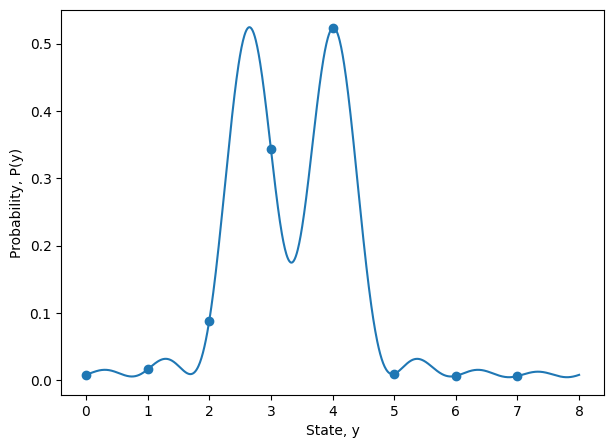}
    \end{center}
    \caption{Curve-fitted QPE results with 1000000 shots and 3 recording qubits. The actual phases were \(\frac{1}{3}\) and \(\frac{1}{2}\). The estimated phases were \(0.33326865\) and \(.50001139\).} 
    \label{fig:examples_2phases}
\end{figure}

\subsection{Comparison with CRLB}\label{subsection:crlb_comparison}

\begin{figure}[ht]
    \begin{center}
    \setkeys{Gin}{width=\linewidth}
    \includegraphics{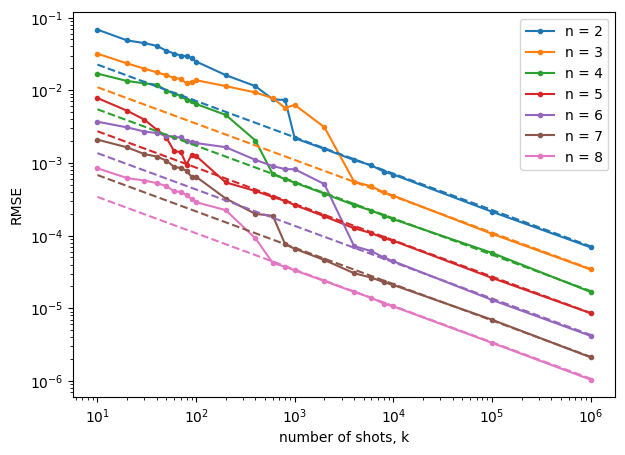}
    \end{center}
    \caption{RMSE against \(k\) with the respective CRLB comparison (dashed lines)}
    \label{fig:against_num_shots_CRLB}
\end{figure}

\begin{figure}[ht]
    \begin{center}
    \setkeys{Gin}{width=\linewidth}
    \includegraphics{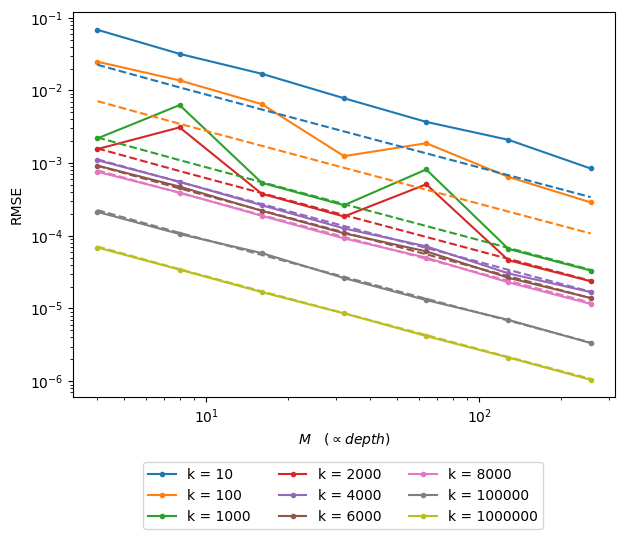}
    \end{center}
    \caption{RMSE against M with the CRLB comparison}
    \label{fig:against_depth_CRLB}
\end{figure}

We compare our results to the CRLB in figures \ref{fig:against_num_shots_CRLB} and \ref{fig:against_depth_CRLB} We see our method achieves consistently optimal performance on the CRLB whenever \(k\) is at least 4000.

\begin{figure}[ht]
    \begin{center}
    \setkeys{Gin}{width=\linewidth}
    \includegraphics{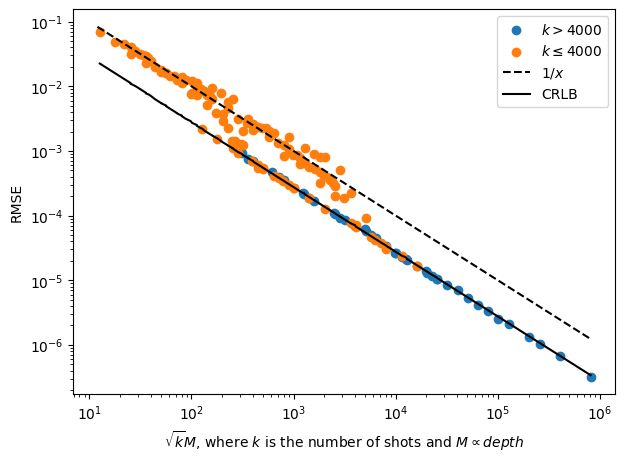}
    \end{center}
    \caption{RMSE, clustered, compared with Heisenberg's Limit and with CRLB comparison (solid line). The dashed line is an approximation made by the authors.}
    \label{fig:against_HL_cluster}
\end{figure}

We can also plot our RMSE against \((\sqrt{k} M)\) in figure \ref{fig:against_HL_cluster} to see that it follows the Heisenberg's Limit. In agreement with the above, our RMSE switches to complete CRLB performance when \(k\) is at least 4000. 

This means that our method can achieve the lowest possible mean squared error due to the fundamental limits of quantum metrology and quantum mechanics.

\section{Conclusion and Further Work} \label{section:conclusion}

We have presented a feasible, working NISQ-era variant of the QPE, which achieves the same error scaling as currently front-running NISQ-era quantum algorithms. We have also shown that our method can achieve optimal CRLB performance. 

Our method consistently produces an advantage in the absolute value precision compared to the traditional QPE without additional quantum resources. Therefore, it is a useful post-processing step to improve the traditional QPE algorithm's precision.

We postulate that our method can simulate or extend current quantum circuits to achieve a sufficiently precise Quantum Phase Estimation result for testing or use in other quantum algorithms. Current quantum computers are still severely limited in size, limiting the precision possible with the traditional QPE method. Our method can extend this precision, following an optimal classical error resolution order demonstrated in section \ref{subsection:resolution_order}, to attain a usable result.

Further work is needed to compare our method to QPE alternatives or variants such as VQE and investigate its advantages and disadvantages. Another possible future area to explore is its potential to find multiple phases or phases when no eigenstate of the system is known. Additionally, our method can be further investigated by implementing a larger quantum algorithm, such as Shor's Algorithm or the HHL Algorithm, to test its application.

\section*{Acknowledgement}

The authors would like to thank QWorld and Classiq Tehnologies for their support, advice, comments and proofreading. R. C. would like to thank E. Cornfeld, O. Samimi, T. Goldfriend and E. Schirman for the fruitful discussion about phase estimation. C. S. acknowledges the University of Córdoba for partial support (Grant N. FCB-PS01-22).

\end{sloppypar}

\bibliography{references}

\appendix
\onecolumngrid

\section{Deriving Probability Mass Function of QPE}
\label{app:pmf}
In this section we will derive the probability equation of the single-eigenphase QPE algorithm step-by-step. A similar derivation can be found in \cite{nielsen2010quantum}.

First, we start with an initial register of \(\vert 0 \rangle^n \otimes \vert \psi \rangle\) and apply a Hadamard on the \(n\)-th qubit. Then we apply \(2^{0}\) times of the Controlled-U gate on the second register of state \(\vert \psi \rangle\), with the \(n\)-th qubit in the first register as the control qubit.

\begin{equation}
CU^{2^0} \left(\frac{1}{\sqrt{2}}(\vert0\rangle + \vert1\rangle) \right) \vert\psi\rangle 
= \frac{1}{\sqrt{2}}\left(\vert0\rangle \vert\psi\rangle+ e^{ 2\pi i {\theta} {2^0}} \vert1\rangle \vert\psi\rangle\right)
=\frac{1}{\sqrt{2}}\left(\vert0\rangle + e^{ 2\pi i {\theta} {2^0}} \vert1\rangle \right)\vert\psi\rangle
\end{equation}

Then we apply a Hadamard on the \(n-1\)-th qubit, and apply the same Controlled-U gate but with the \(n-1\)-th qubit as the control qubit, and we apply the Controlled-U gate \(2^{1}\) times instead, to get

\begin{equation}
\frac{1}{\sqrt{4}}\left(\vert0\rangle + e^{2\pi i {\theta} {2^1}} \vert1\rangle \right)\left(\vert0\rangle + e^{2\pi i {\theta} {2^0}} \vert1\rangle \right)\vert\psi\rangle
\end{equation}

We repeat this with all the remaining \(n-2\) qubits, to get
\begin{equation}
\frac{1}{\sqrt{M}}\left(\vert0\rangle + e^{2\pi i {\theta} {2^{n-1}}} \vert1\rangle \right)\left(\vert0\rangle + e^{2\pi i {\theta} {2^{n-2}}} \vert1\rangle \right)...\left(\vert0\rangle + e^{2\pi i {\theta} {2^0}} \vert1\rangle \right)\vert\psi\rangle
\end{equation}
where \(M = 2^n\).

This equals to 
\begin{equation}
\vert \phi_1 \rangle = \frac{1}{\sqrt{M}}\sum_{x=0}^{M-1}e^{ 2\pi i \theta x} \vert x \rangle \vert\psi\rangle
\end{equation}

Noting that the inverse Quantum Fourier Transform results in the following transformation, 
\begin{equation}
{QFT}^{-1}\vert x \rangle = \frac{1}{\sqrt{M}}
\sum^{M-1}_{y=0} e^{\frac{-2\pi i yx}{M}} \vert y \rangle
\end{equation}

We apply the inverse QFT on our current qubits \(\vert x \rangle\) to get
\begin{equation}
\begin{split}
\vert \phi_2 \rangle & = \frac{1}{\sqrt{M}}\sum_{x=0}^{M-1}e^{ i 2 \pi \theta x} \left( \frac{1}{\sqrt{M}} \sum^{M-1}_{y=0} e^{\frac{-2\pi i yx}{M}} \vert y \rangle \right) \vert\psi\rangle \\
& = \frac{1}{M} \sum_{y=0}^{M-1} \sum^{M-1}_{x=0} e^{\frac{-2\pi i yx}{M}} e^{ i 2 \pi \theta x} \vert y \rangle \vert\psi\rangle \\
& = \frac{1}{M} \sum_{y=0}^{M-1} \sum^{M-1}_{x=0} e^{\frac{-2\pi i x}{M}\left(y - \theta M \right)} \vert y \rangle \vert\psi\rangle \\
\end{split}
\end{equation}
This is our final state after applying QPE on the basic case.

We see that the probability to measure any \(\vert y \rangle\) state is 
\begin{equation}
P(y) = \left| \frac{1}{M}\sum^{M-1}_{x=0} e^{-\frac{2\pi i x}{M}\left({y - \theta M}\right)} \right| ^2
\end{equation}

Using the equation for the sum of a geometric progression series, 

\begin{equation}
\begin{split}
P(y)  & = \left| \frac{1}{M} \sum^{M-1}_{x=0} e^{-\frac{2\pi i x}{M}\left(y - \theta M\right)} \right| ^2 \\
& = \frac{1}{M^2} \left| \frac{1 - e^{-\frac{2\pi i}{M}\left(y - \theta M\right)M}}{1 - e^{-\frac{2\pi i}{M}\left(y - \theta M\right)}} \right| ^2 \\
& = \frac{1}{M^2} \left| \frac{e^{-\frac{\pi i}{M}\left(y - \theta M\right)M}}{e^{-\frac{\pi i}{M}\left(y - \theta M \right)}} \right| ^ 2 \left| \frac{e^{\frac{\pi i}{M}\left(y - \theta M \right)M} - e^{-\frac{\pi i}{M}\left(y - \theta M \right)M}}{e^{\frac{\pi i}{M}\left(y - \theta M \right)} - e^{-\frac{\pi i}{M}\left(y - \theta M\right)}} \right| ^ 2 \\
& = \frac{1}{M^2} \left| \frac{2 i \sin(\frac{\pi }{M}\left(y - \theta M \right)M)}{2 i \sin(\frac{\pi }{M}\left(y - \theta M \right))} \right| ^ 2\\
& = \frac{1}{M^2} \left( \frac{ \sin^2(\frac{\pi }{M}\left(y - \theta M \right)M)}{ \sin^2(\frac{\pi }{M}\left(y - \theta M\right)} \right)\\
& = \frac{1}{M^2} \left( \frac{ 1 - \cos(\frac{2\pi }{M}\left(y - \theta M\right)M)}{ 1 - \cos(\frac{2\pi }{M}\left(y - \theta M \right))} \right)\\
& = \frac{1}{M^2} \left( \frac{ 1 - \cos(2\pi\left(y - \theta M \right))}{ 1 - \cos(\frac{2\pi }{M}\left(y - \theta M \right))} \right)\\
\end{split}
\end{equation}

\section{Deriving Generic Probability Mass Function of QPE}
\label{app:generic_pmf}

For the case where \(\bm{U} \sum a_j \vert \psi_j\rangle = \sum a_j e^{ i {\theta_j}} \vert {\psi_j} \rangle\), we let \(\vert \psi \rangle = \sum a_j \vert \psi_j\rangle\).

After applying Hadamard gates to the \(n\) qubits, we have the state

\begin{equation*}
\vert \phi_1 \rangle = \frac{1}{\sqrt{M}}\sum_{x=0}^{M-1} \vert x \rangle \vert\psi\rangle
\end{equation*}

We then apply the Controlled-U gates,

\begin{equation*}
\begin{split}
\vert \phi_2 \rangle & = \frac{1}{\sqrt{M}}\sum_{x=0}^{M-1} \vert x \rangle U^x \vert\psi\rangle \\
& = \frac{1}{\sqrt{M}}\sum_{x=0}^{M-1} \vert x \rangle \sum_j a_j e^{ 2 \pi i {\theta_j} x} \vert {\psi_j} \rangle \\
\end{split}
\end{equation*}

We then apply the inverse QFT on \(\vert x \rangle\),
\begin{equation}
\begin{split}
\vert \phi_3 \rangle & = \frac{1}{\sqrt{M}}\sum_{x=0}^{M-1} \left(\frac{1}{\sqrt{M}} \sum^{M-1}_{y=0} e^{\frac{-2\pi i yx}{M}} \vert y \rangle \right) \sum_j a_j e^{ 2\pi i {\theta_j} x} \vert {\psi_j} \rangle \\
& = \frac{1}{M} \sum^{M-1}_{x=0}\sum_{y=0}^{M-1}  \sum_j a_j e^{-\frac{2\pi i x}{M} \left({y - \theta_j M} \right)} \vert y \rangle \vert {\psi_j} \rangle \\
\end{split}
\end{equation}

Hence, the probability of measuring a state \(y\) is given by
\begin{equation}
\begin{split}
P & = \left| \frac{1}{M} \sum_{x=0}^{M-1}  \sum_j a_j e^{-\frac{2\pi i x}{M} \left({y - \theta_j M} \right)} \right| ^ 2 \\
& = \frac{1}{M^ 2} \sum_j a_j^2 \left|  \frac{ \left( 1 - e^{-\frac{2\pi i }{M} \left({y - \theta_j M} \right) M }\right) }{\left( 1 - e^{-\frac{2\pi i }{M} \left({y - \theta_j M} \right)}\right)} \right| ^ 2 \\
& = \frac{1}{M^ 2} \sum_j a_j^2 \left| \frac{e^{-\frac{\pi i }{M} \left({y - \theta_j M} \right) M}}{e^{-\frac{\pi i }{M} \left({y - \theta_j M} \right)}}\right| ^ 2 \left| \frac{ \left( e^{\frac{\pi i }{M} \left({y - \theta_j M} \right) M} - e^{-\frac{\pi i }{M} \left({y - \theta_j M} \right) M}\right) }{\left(e^{\frac{\pi i }{M} \left({y - \theta_j M} \right)} - e^{-\frac{\pi i }{M} \left({y - \theta_j M} \right)}\right)} \right| ^ 2 \\
& = \frac{1}{M^ 2} \sum_j a_j^2  \left| \frac{  2i \sin\left({\frac{\pi  }{M} \left({y - \theta_j M} \right) M}\right) }{2i\sin\left({\frac{\pi  }{M} \left({y - \theta_j M} \right)}\right)} \right| ^ 2 \\
& = \frac{1}{M^ 2} \sum_j a_j^2  \left( \frac{ \sin^ 2\left({\frac{\pi  }{M} \left({y - \theta_j M} \right) M}\right) }{\sin^ 2\left({\frac{\pi  }{M} \left({y - \theta_j M} \right)}\right)} \right) \\
& = \frac{1}{M^ 2} \sum_j a_j^2  \left( \frac{ 1-\cos\left({2\pi \left({y - \theta_j M} \right) }\right) }{1-\cos\left({\frac{2\pi  }{M} \left({y - \theta_j M} \right)}\right)} \right) \\
\end{split}
\end{equation}

\section{Fisher Information Values}
\label{app:fisher}
The Fisher Information value for an individual shot for each choice of M is given below 

\begin{center} \label{table:fi}
\begin{tabular}{||c c||} 
 \hline
 M & FI \\ [0.5ex] 
 \hline\hline
 4 & 197.39208802 \\ 
 \hline
 8 & 829.04676969 \\
 \hline
 16 & 3355.66549637 \\
 \hline
 32 & 13462.14040308 \\
 \hline
 64 & 53888.04002995 \\ 
 \hline
 128 & 215591.6385373 \\ 
 \hline
 256 & 862406.03256634 \\ [1ex] 
 \hline
\end{tabular}
\end{center}

\end{document}